# Time-resolved study of the magnetic field effects on electroluminescence in tri-(8-hydroxyquinoline)- aluminum based organic light emitting devices

Qiming Peng, Xianjie Li, Feng Li[*]

State Key Lab of Supramolecular Structure and Materials, Jilin University, 2699

Qianjin Avenue, Changchun 130012, People's Republic of China

**Abstract**: We investigated the magnetic field effects (MFEs) in organic light-emitting diodes (OLEDs) through the transient electroluminescence (EL) method. The time-resolved MFEs on the emission were obtained for the first time, which would be a useful method to clarify the underlying mechanisms of the MFEs. The fluorescent dye doped tri-(8-hydroxyquinoline)-aluminum ($Alq_3$) based OLEDs were fabricated. Then, the transient EL was measured both with and without a magnetic field. To explore the time-resolved MFEs on the emission of the device, the excitons population dynamics in the device have been analyzed by a kinetic model. Our results suggest that both the intersystem crossing between the singlet and triplet electron-hole pairs and the triplet-triplet annihilation perturbed by the external magnetic field cause the time-resolved MFEs.

*Author to whom correspondence should be addressed; Electronic mail:
   lifeng01@jlu.edu.cn

Since kalinowski *et al.* reported that the current and electroluminescence (EL) of tri-(8-hydroxyquinoline)-aluminum (Alq$_3$) based organic light-emitting devices (OLEDs) can be enhanced by a relatively weak magnetic field (<500mT) [1], the magnetic field effects (MFEs) on the electroluminescence (namely, MEL) and on the current (namely, MC) of the organic devices with non-magnetic electrodes have been investigated extensively [2-25]. To explore these novel phenomena, several models have been proposed, such as the electron-hole pairs (EHP) model [1,8], the bipolaron model [5], the triplet-polaron interaction (TPI) model [9.15], the triplet-triplet annihilation (TTA) model [16,17], and so on. It should be mentioned that even though each of these models can account for some experimental phenomena, none of them are suited for a fully consistent interpretation of all data reported. Therefore, further studies might find ways to explain the MFEs by combining different models [1,5,9,13,16] since each model might be the major mechanism under only a particular condition. In fact, most of the studies reported earlier were focused on the steady-state investigation (in which OLEDs were driven by a constant voltage or current source). However, the transient method (in which OLEDs were driven by a pulse voltage) has not been paid much attention in the MFEs investigation, even though it could act as a significant method to investigate the charge kinetics.

The transient method has been used intensively for decades to study the charge processes in organic semiconductors such as charge injection, transport, recombination, and excitons decay [18-26]. Barth *et al.* and other groups have studied the carriers mobility by measuring the onset time of the EL pulse of the OLEDs driven by a rectangular pulse voltage [18,21]. Luo *et al.* have studied the delay fluorescence by analyzing the tail of the transient EL signals [22,23]. Indeed, the transient method has some superiorities compared to the steady-state method, due to the fact that different charge processes occur in different periods of the pulse. Whether or not these processes are magnetic field dependent could be investigated by measuring the transient signals when the magnetic field is either on or off. In our previous work [24], we have investigated the MFEs on the carriers' mobility and the bimolecular recombination by the transient method. The purpose of this paper is to investigate the MFEs on the transient EL and analyze the time-resolved MELs. The time-resolved MEL was obtained by $MEL(t)=\Delta EL(t)/EL(t,0)=[EL(t,B)-EL(t,0)]/EL(t,0)$ from the EL curves with and without the magnetic field. Furthermore, we provide a model to detail the charge processes and theoretically simulate the time-resolved MELs data.

In this work, $Alq_3$ based devices with the structure (see in the insert (c) of figure 1) of indium tin oxide (ITO)/*N*,*N*-di-1-naphthyl-*N*,*N*-diphenylbenzidine (NPB 50 nm)/ $Alq_3$ : 10-(2-Benzothiazolyl)-2,3,6,7- tetrahydro-1,1,7,7-tetramethyl-1H,5H,11H-(1) benzopyropyrano ( 6, 7-8-I, j ) quinolizin -11-one (C545T) (5 wt. %, 30 nm)/ $Alq_3$ (40nm) /LiF (0.8nm)/Al (100nm) were fabricated by using the multiple-source organic molecular beam deposition method. Immediately after fabrication, the devices were placed on a Teflon stage between the poles of an electromagnet for the MELs measurement. We used an Agilent 8114A pulse generator (100 v/2 A) to apply rectangular pulse voltages to our devices. The pulse repetition rate was 1 KHz with a width of 4 μs. The light output of the devices was collected by a lens coupled with the optical fiber (2 m) connected to a Hamamatsu photomultiplier (H5783P–01, time resolution: 0.78 ns). The photomultiplier was placed far away from the electromagnet and was connected to one of the channel of a digital oscilloscope (Tektronix DPO7104, sampling rate: 5 GS/s; resolution:100 μV) with a 50 Ω input resistance. All measurements were carried out at room temperature under ambient condition.

Figure 1 shows a set of transient EL signals of the device driven by the pulse voltages ranging from 5 V to 15 V with the magnetic field of 150 mT being on (red line) and off (black line). We can see that the final overshoots of the EL pulses can be seen at all voltages (see the inserts (a) and (b) of figure 1), while the initial overshoots are only observed at the voltages above 11 V and become remarkable as the voltage increases. The values of MELs are positive at the voltages below 11 V (see the insert (b) of figure 1) and turn from positive at the initial overshoot region to negative at the flat region (see the insert (a) of figure 1) with the time at the voltages above 11 V.

Figure 2 shows the time-resolved MELs with various driving pulse voltages (6 V, 10 V and 15 V). It can be seen that the MEL curves show similar behaviors at the pulse-off period ($t > 4$ μs). A sudden rise followed by saturation appears for all driving voltages. However, at the pulse-duration period ($t = 0~4$ μs), the MEL curves show very different behaviors. In this period the MELs decrease with the time and the decreasing slope is proportional to the driving voltage. Negative MELs are seen when the driving voltage is high enough ($V > 11$V in our experiment).

To analyze the time-resolved MELs, the charge carriers and excitons population dynamics in the devices driven by the pulse voltages need to be considered. Generally, the excitons population dynamics in organic semiconductor devices involve triplet-triplet annihilation (TTA), triplet-polaron interaction (TPI), excitons dissociation (ED), singlet-polaron interaction (SPI), radiate decay (RD) and non-radiate decay (NRD) of the excitons, intersystem crossing (ISC), and the triplet-singlet interaction (TSI) etc.. Figure 3 shows the schematic diagram of the charge and excitons kinetics in the emitting layer. For simplicity, the diffusion process in the interface and other faint interactions such as SPI and triplet dissociation have been ignored.

Herein, we provide a model based on the bimolecular recombination to analyze the time-resolved MELs. When the rectangle pulse voltage arrives, the charge carriers are injected into the electrode, drift in the transport layer and fill up the traps, then recombine with the opposite-charge carriers to build singlets and triplets with the ratio of 1:3 according to the spin statistics rule [6]. The population dynamic of the free charge carriers could be expressed as:

$$\frac{dn}{dt} = \frac{J(t)}{ed} - \gamma n^2 \tag{1}$$

Where $n$ is the density of the free charge carriers (electrons or holes), $J(t)$ is the injection current density as a function of time, $e=1.6*10^{\wedge}(-19)$ is the charge of the carriers, $d$ represent the width of the recombination region, and $\gamma$ is the recombination rate [27].

The triplet and singlet population dynamics in the device involve TTA, TPI, RD, NRD, TSI and ISC etc.. These processes could be expressed as:

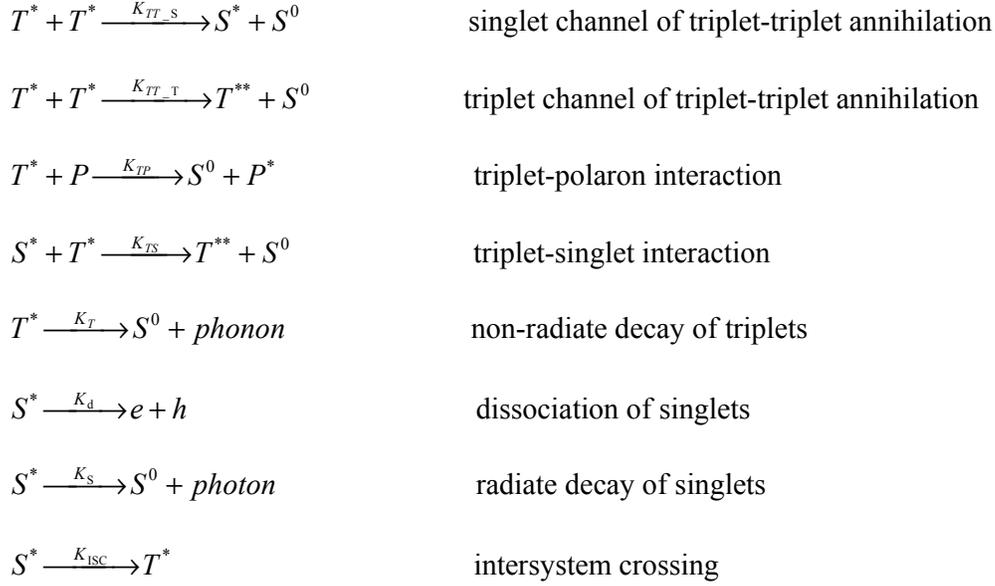

$$T^* + T^* \xrightarrow{K_{TT\_S}} S^* + S^0 \qquad\qquad \text{singlet channel of triplet-triplet annihilation}$$

$$T^* + T^* \xrightarrow{K_{TT\_T}} T^{**} + S^0 \qquad\qquad \text{triplet channel of triplet-triplet annihilation}$$

$$T^* + P \xrightarrow{K_{TP}} S^0 + P^* \qquad\qquad \text{triplet-polaron interaction}$$

$$S^* + T^* \xrightarrow{K_{TS}} T^{**} + S^0 \qquad\qquad \text{triplet-singlet interaction}$$

$$T^* \xrightarrow{K_T} S^0 + phonon \qquad\qquad \text{non-radiate decay of triplets}$$

$$S^* \xrightarrow{K_d} e + h \qquad\qquad \text{dissociation of singlets}$$

$$S^* \xrightarrow{K_S} S^0 + photon \qquad\qquad \text{radiate decay of singlets}$$

$$S^* \xrightarrow{K_{isc}} T^* \qquad\qquad \text{intersystem crossing}$$

In consideration of the excitons kinetics above, the triplet density could be written as:

$$\frac{dT}{dt} = \frac{3}{4}\gamma n^2 + K_{isc}S - K_T T - (K_{TT\_T} + 2K_{TT\_S})T^2 - K_{TP}nT \qquad (2)$$

Where $T$ is the triplet density and $S$ represents the singlet density, $K_{isc}$ is the ISC rate, $K_T$ is the NRD rate of the triplet, and $K_{TT\_T}$ and $K_{TT\_S}$ represent the *TTA* rate of the singlet channel and triplet channel, respectively. The singlet channel of the TTA must satisfy the spin conservation requirement; that is, only two processes, $T^*_{m=0} + T^*_{m=0} = S^* + S^0$, and $T^*_{m=+1} + T^*_{m=-1} = S^* + S^0$, occur in the singlet channel [6], and the next four processes of TTA occur in the triplet channel, which leads to the ratio of the $K_{TT\_S}$ to $K_{TT\_T}$ as 1:2. $K_{TP}$ is the TPI rate. The constant 3/4 is controlled by the spin statistics rule.

The singlet dynamic equation can be written as:

$$\frac{dS}{dt} = \frac{1}{4}\gamma n^2 - (K_{isc} + K_d + K_s)S - K_{TS}TS + K_{TT\_S}T^2 \qquad (3)$$

Where $K_S$ represents the singlet RD rate, $K_d$ is the dissociation rate of singlet which can be affected by the electrical field, and $K_{TS}$ represents the TSI rate. Finally, the constant 1/4 is dependent on the spin statistics rule.

Given the ordinary differential equations (1), (2), and (3), we can simulate the singlet and triplet population dynamics. It has to be mentioned that the parameter $K_a = (K_{isc} + K_d + K_s)$ was used in this simulation. Some of the rate constants in this simulation were obtained from the literature data for the typical doped systems [27-29]. According to the published works, the action rate constants in this simulation are magnetic field independent, except for $K_{isc}, K_{TT\_S}$, and $K_{TP}$. The MFEs on the $K_{isc}$ were suggested by Kalinowski *et al.* [1]. Johnson and Merrifield *et al.* suggested that the singlet channel of the TTA could be affected by an external magnetic field [30-32]. Gillin *et. al.* proposed that an external magnetic field could affect the $K_{TP}$ [9]. However, our previous work demonstrated that the TPI model could not be a major mechanism of the MEL in our experiment conditions [33], thus only the rates $K_{isc}$ and $K_{TT\_S}$ are functions of the external magnetic field in the simulation. By setting $n=0$, $S=0$, and $T=0$ at zero-time (corresponding to the time of EL onset, $t_{EL\_onset}$) as the initial conditions of the differential equations and adjusting the $K_a$ and $K_{TT\_S}$ to different values for the magnetic field being on and off, we simultaneously obtain the numerical solutions of the ordinary differential equations (1), (2) and (3). The calculated triplet and singlet densities for the pulse-duration period are shown in figure 4 (a) and figure 4 (b), respectively. The black line and red line represent the calculated results with the magnetic field being on and off, respectively. Particularly, the time-resolved singlet density could be regarded as the time-resolved EL because only the singlet could radiately decay in the fluorescent materials according to the spin select rule.

From figure 4 (a), we can see that the triplet density is small at the beginning of the pulse-duration period. Because the TTA is quadratic proportional to the triplet density [6], the TTA process at this period can be ignored and has little contribution to the total MEL. The total MEL at the beginning of the pulse-duration period is attributed to the change of the $K_{isc}$ ($K_a$ in this paper). When the external magnetic field is present and larger than the hyperfine field, the degeneracy of the triplet substates is removed, leading to the reduced mixing between the singlet e-h pairs and triplet e-h pairs; thus, the singlet population increases [1], as shown in figure 4 (b). As the time increases, the triplet density increases, and the TTA process must be considered. The singlet channel of TTA could be suppressed by a high external magnetic field [32], leading to an increase of the triplet population and a decrease of the singlet population, as shown in figure 4 (a) and (b), respectively. With the increase of the triplet density, another unavoidable effect is the triplet-singlet interaction, which could induce the quenching of the singlet excitons [28,29]. This is the reason that the initial overshoot exists at the beginning of the EL pulse. Figure 4 (c) shows the experimental results. We can see that our simulation results (figure 4 (b)) coincide perfectly with the experimental data. During the simulation, we solely adjusted the $K_a(B)$ and $K_{TT\_S}(B)$ for both with and without the magnetic field. It turns out that the changing ratio of $K_{TT\_S}(B)$ (- 5%) is very close to the value calculated from literature data [17,34]. All the values of the parameters used in this simulation are shown in figure 4.

By the same process, we obtained the simulation results for the driving pulse voltage of 6 V and 10 V. Figure 5 shows the simulated time-resolved MELs together with the experimental data for various driving voltages (6V, 10V and 15V). The black line and red line represent the experimental and simulation results, respectively. As can be seen, they are consistent with each other. We found that combining the magnetic-field-modified ISC (positive effect) and TTA (negative effect) could ideally explain the time-resolved MELs. At the beginning of the pulse-duration period, because the triplet density is small, the magnetic-field-modified ISC is dominant. As the time increases, the density of triplet excitons increases. The magnetic-field-modified TTA becomes un-neglectable. Thus the total MEL declines with the time. At higher driving voltage, the density of triplet excitons increase more quickly with the time, which induces the decreasing slope of the MEL to become steeper. When the magnetic-field-modified TTA is dominant, the negative MEL can be seen.

For the pulse-off period, it is unsuitable to analyze the time-resolved MELs by using the model provided above, because the pulse-off period involves trapped charge carriers and their detrapping process [19]. This will be analyzed elsewhere.

Recently, Koopmans *et al.* have investigated the dependence of MC on the frequency of the applied magnetic field, which consists of a dc and ac component [35,36]. We have studied the time-resolved MEL by applying the pulse voltages. One focuses on the carriers' transporting process, the other focuses on the carriers' recombination process. Thus the two methods are very complementary to each other.

In summary, we provided an alternative way (time-resolved MFEs) to study the magnetic field effects of the OLEDs. To analyze the time-resolved MELs, we modeled the charge population dynamics. The simulation results coincide well with the experimental results. It is found that both the magnetic-field-modified ISC and TTA result in the time-resolved MELs. The method of time-resolved MFEs has strong potential for studying the underlying mechanisms of the MFEs in organic semiconductor devices.

We are grateful for financial support from the National Natural Science Foundation of China (grant numbers 60878013).

**References**

[1] J. Kalinowski *et al.*, Chem. Phys. Lett. **380**, 710 (2003).

[2] B. Hu and Y. Wu, Nat. Mater. **6**, 985 (2007).

[3] F. L. Bloom *et al.*, Phys. Rev. Lett. **99**, 257201 (2007).

[4] J. D. Bergeson *et al.*, Phys. Rev. Lett. **100**, 067201 (2008).

[5] P. A. Bobbert *et al.*, Phys. Rev. Lett. **99**, 216801 (2007).

[6] B. Hu *et al.*, Adv. Mater. **21**, 1500 (2009).

[7] B. Ding *et al.*, Phys. Rev. B **82**, 205209 (2010).

[8] S. A. Bagnich *et al.*, J. Appl. Phys. **105**, 123706 (2009).

[9] P. Desai *et al.*, Phys. Rev. B **75**, 094423 (2007).

[10] N. Rolfe *et al.*, J. Appl. Phys. **104**, 083703 (2008).

[11] F. L. Bloom *et al.*, Phys. Rev. Lett. **103**, 066601 (2009).

[12] T. D. Nguyen *et al.*, Phys. Rev. B **77**, 235209 (2008).

[13] F. J. Wang *et al.*, Phys. Rev. Lett. **101**, 236805 (2008).

[14] Y. B. Zhang *et al.*, Org. Electron. **9**, 687 (2008).

[15] J. Song *et al.*, Phys. Rev. B **82**, 085205 (2010).

[16] A. H. Davis and K. Bussmann, J. Vac. Sci. Technol. A **22**, 1885 (2004).

[17] P. Chen *et al.*, Appl. Phys. Lett. **95**, 213304 (2009).

[18] V. Savvateev *et al.*, Adv. Mater. **11**, 519 (1999).

[19] N. D. Nguyen *et al.*, Phys. Rev. B **75**, 075307 (2007).

[20] J. Kalinowski *et al.*, Appl. Phys. Lett. **72**, 513 (1998).

[21] S. Barth *et al.*, J. Appl. Phys. **89**, 3711 (2001).

[22] Y. Luo and H. Aziz, Adv. Funct. Mater. **20**, 1285 (2010).

[23] Y. Luo and H. Aziz, J. Appl. Phys. **107**, 094510 (2010).

[24] F. Li *et al.*, Appl. Phys. Lett. **97**, 073301 (2010).

[25] L. Xin *et al.*, Appl. Phys. Lett. **95**, 123306 (2009).

[26] L. Hassine *et al.*, Appl. Phys. Lett. **78**, 1053 (2001).

[27] N. C. Giebink and S. R. Forrest, Phys. Rev. B **79**, 073302 (2009).

[28] Y. Zhang *et al.*, Chem. Phys. Lett. **495**, 161 (2010).

[29] C. Ga rtner *et al.*, J. Appl. Phys. **101**, 023107 (2007).

[30] R. C. Johnson *et al.*, Phys. Rev. Lett. **19**, 285 (1967).

[31] V. Ern and R. E. Merrifield, Phys. Rev. Lett. **21**, 609 (1968).

[32] R. C. Johnson and R. E. Merrifield, Phys. Rev. B **1**, 896 (1970).

[33] Q. Peng *et al.*, Appl. Phys. Lett. accepted

[34] R. Liu *et al.*, J. Appl. Phys. **105** , 093719 (2009).

[35] W. Wagemans *et al.*, Appl. Phys. Lett. **97**, 123301 (2010).

[36] P. Janssen *et al.*, Synth. Met. **161**, 617 (2011).

**Figure captions**

**FIG.1**. The transient EL responses of the device driven by various rectangular pulse voltages (repetition: 1 KHz; width: 4 μs) with the magnetic field (150 mT) being on (red line) and off (black line). The inserts (a) and (b) show the enlargement of the EL curves at the pulse-duration period when the driving voltages are 7 V and 15 V, respectively. The insert (c) is a scheme of the structure of the device.

**FIG. 2**. The time-resolved MELs (calculated by $MEL=[EL(t,B)-EL(t,0)]/EL(t,0)$) of the device driven by various rectangular pulse (6 V, 10 V and 15 V).

**FIG. 3**. Schematic diagram for carriers and excitons kinetics in the emitting layer.

**FIG. 4**. The simulation results of the triplet (a) and singlet (b) density as a function of time with the magnetic field being on (red line) and off (black line) for the driving voltage of 15 V, all the values of the parameters used in the simulation are shown beside. (c) the experimental data of the EL response driven by a voltage of 15 V with the magnetic field of 150 mT turned on (red line) and off (black line).

**FIG. 5**. The simulated time-resolved MELs (red line) and the experimental data (black line) at the pulse-duration period for various driving voltages (6 V, 10 V, and 15 V). The onset time of the EL ($t_{EL\_onset}$) and the corresponding currents density in the device at each voltage are shown in each part of the figure.

**FIG.1**

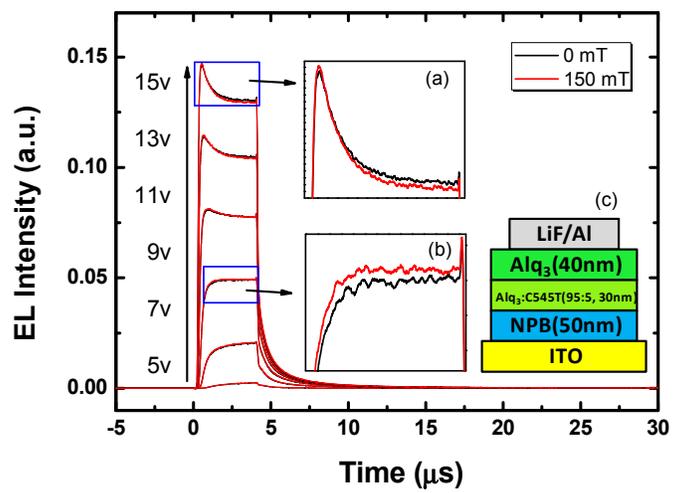

**FIG. 2**

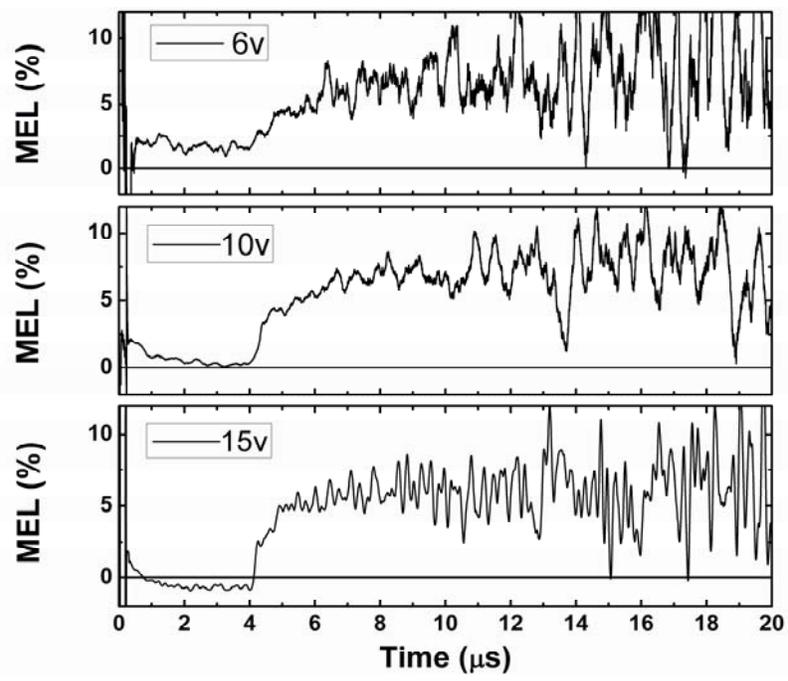

**FIG. 3**

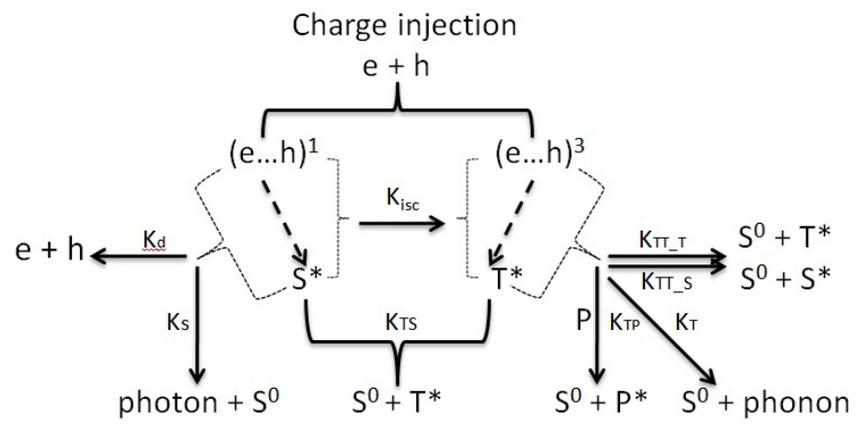

**FIG. 4**

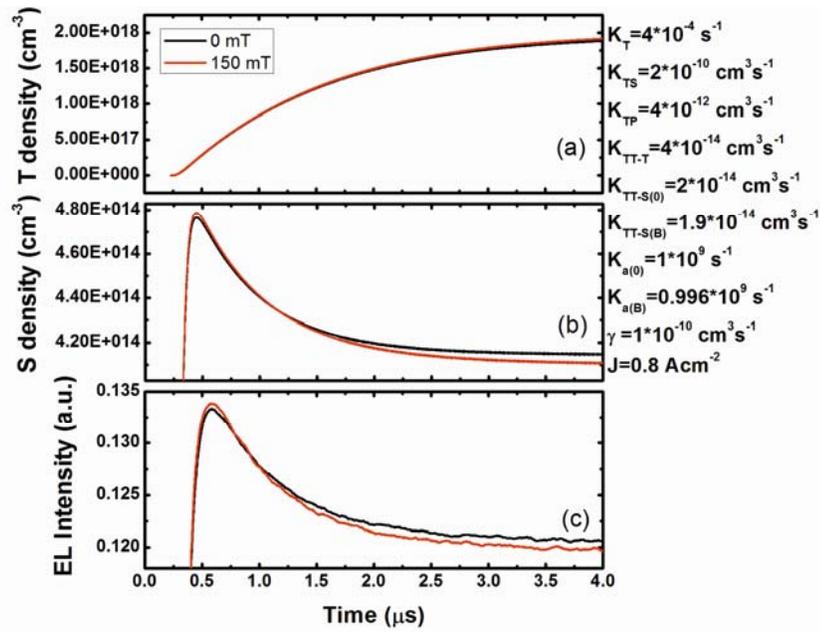

The plot panels include the following parameters:

$K_T = 4 \times 10^{-4}\ s^{-1}$
$K_{TS} = 2 \times 10^{-10}\ cm^3 s^{-1}$
$K_{TP} = 4 \times 10^{-12}\ cm^3 s^{-1}$
$K_{TT\text{-}T} = 4 \times 10^{-14}\ cm^3 s^{-1}$
$K_{TT\text{-}S(0)} = 2 \times 10^{-14}\ cm^3 s^{-1}$
$K_{TT\text{-}S(B)} = 1.9 \times 10^{14}\ cm^3 s^{-1}$
$K_{a(0)} = 1 \times 10^9\ s^{-1}$
$K_{a(B)} = 0.996 \times 10^9\ s^{-1}$
$\gamma = 1 \times 10^{-10}\ cm^3 s^{-1}$
$J = 0.8\ Acm^{-2}$

**FIG. 5**

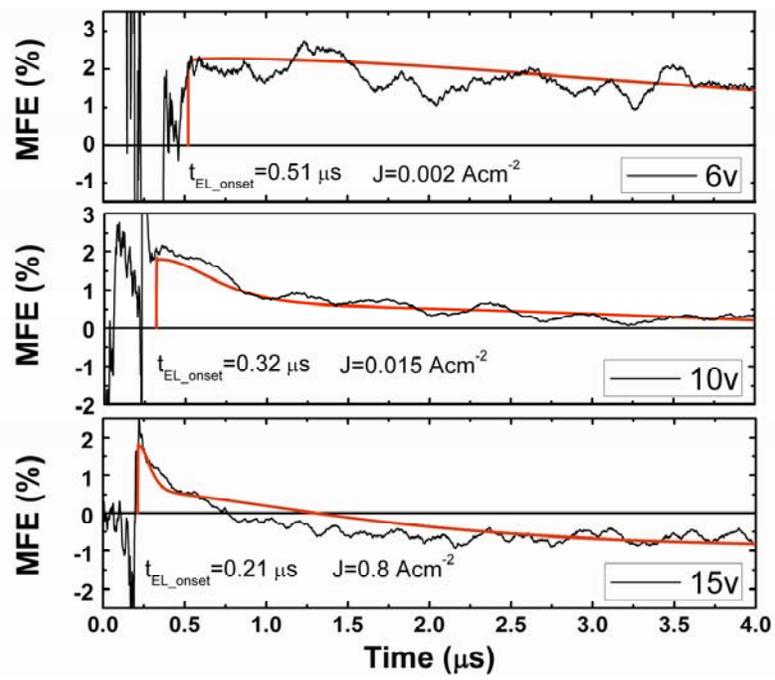